\documentclass[10pt,review]{elsarticle}

\usepackage{graphicx}

\usepackage{url}

\begin{document}
\begin{frontmatter}
\title{High-Dimensional Data Visualization by Interactive Construction of Low-Dimensional Parallel Coordinate Plots}

\author[authora]{Takayuki~Itoh}\ead{itot@is.ocha.ac.jp}
\author[authorb]{Ashnil~Kumar}\ead{ashnil.kumar@sydney.edu.au}
\author[authorc]{Karsten~Klein}\ead{karsten.klein@monash.edu}
\author[authorb]{Jinman~Kim}\ead{jinman.kim@sydney.edu.au}%

\address[authora]{Ochanomizu University.}
\address[authorb]{The University of Sydney.}
\address[authorc]{Monash University.}

\begin{abstract}
Parallel coordinate plots (PCPs) are among the most useful techniques for the visualization and exploration of high-dimensional data spaces. They are especially useful for the representation of correlations among the dimensions, which identify relationships and interdependencies between variables. However, within these high-dimensional spaces, PCPs face difficulties in displaying the correlation between combinations of dimensions and generally require additional display space as the number of dimensions increases.
In this paper, we present a new technique for high-dimensional data visualization in which a set of low-dimensional PCPs are interactively constructed by sampling user-selected subsets of the high-dimensional data space.
In our technique, we first construct a graph visualization of sets of well-correlated dimensions. Users observe this graph and are able to interactively select the dimensions by sampling from its cliques, thereby dynamically specifying the most relevant lower dimensional data to be used for the construction of focused PCPs. Our interactive sampling overcomes the shortcomings of the PCPs by enabling the visualization of the most meaningful dimensions (i.e., the most relevant information) from high-dimensional spaces.
We demonstrate the effectiveness of our technique through two case studies, where we show that the proposed interactive low-dimensional space constructions were pivotal for visualizing the high-dimensional data and discovering new patterns.
\end{abstract}



\end{frontmatter}

\section{Introduction}
Many data analysis tasks can be facilitated by high-dimensional data visualization.
For example, it is often helpful to know what sets of dimensions in the data are correlated 
or have a significant impact in processes such as clustering, sampling, or labeling.
Another important task is the discovery of hidden relationships between labels and numeric values in the analysis of labeled high-dimensional datasets.
Several machine learning techniques such as deep neural networks or association rule mining are useful for this purpose but may require significant computational time when discovering relationships between labels and large arbitrary combinations of numeric dimensions.
Moreover, in many application domains, high-dimensional data analysis requires interactive analysis and decision making by a domain expert to specify the rules between the labels and the numeric values. Such circumstances arise in important application domains such as
biomedicine, finance, and social media analysis, where the a priori characterization and detection of interesting patterns is difficult and limited due to a lack of domain-specific knowledge as well as the complexity and dynamic nature of the data. 
The application of a data visualization framework is often helpful for these types of problems.

High-dimensional data visualization continues to be an important and active research field with several survey papers dedicated to this area~\cite{Gri01} \cite{Won97}. One of these surveys~\cite{Won97} divided the available multi-dimensional data visualization techniques into three categories: animations, two-variate displays, and multivariate displays. Animation techniques facilitate the dynamic display of multiple configurations of the high-dimensional data and are commonly applied to both two-variate and multivariate techniques.
Two-variate techniques only visualize the relationships between two variables; an example of such a technique is the well-known scatterplot (SP). Multivariate visualization techniques attempt to represent the distribution of all the dimensions in a given dataset on a single display space. The increasing dimensionality of modern datasets is spurring the use of multivariate visualization techniques, such as SP matrices and parallel coordinate plots (PCPs) \cite{Ins90}.

A SP matrix consists of multiple adjacent SPs in a grid-like arrangement, with
each SP being identified by its row and column index.
The greatest advantage of SP matrices is the high level of familiarity that non-expert users have with it. However, this visualization technique suffers from few major drawbacks.
Firstly, individual SPs within the display space may be very small if the number of dimensions of the dataset is very large.
In addition, it can be difficult for humans to visually compare arbitrary pairs of SPs that are distantly placed in the display space.
Several studies aimed at selecting meaningful sets of SPs and effectively arrange them onto the display spaces \cite{Dan14} \cite{Sip09} \cite{Wil05} \cite{Yan07} \cite{Zhe15}; however, these studies often had difficulty to visually compare large number of SPs.

PCPs are an alternative multivariate visualization technique that display high-dimensional datasets as a set of polylines that intersect with parallel axes; this visualization better enables the observation of correlations between pairs of dimensions.
Specifically, the existence of parallel polylines between two axes indicates positive correlation. Conversely, polylines crossing between axes are indicative of negative correlation. 
While PCPs have been well studied for high-dimensional data visualization and are a valuable tool for the analysis of multidimensional data, they have some shortcomings in practice.
PCPs may require a large horizontal display space even for a modest number of dimensions. Within this horizontal display space the ordering of the dimensions is relevant for the detection of pairwise correlations.
Moreover, it is difficult to visually represent the correlation of a particular dimension with three or more different dimensions.
An example of this situation can be illustrated given a dataset which has four dimensions $a$, $b$, $c$, and $d$, which are visualized by a single PCP.
The correlations between $a$ and $b$, or $b$ and $c$, can be easily observed if the dimensions are arranged in the order $a$, $b$, $c$, and $d$.
However, the correlation between $b$ and $d$ cannot be easily observed within this visualization. 

Several techniques \cite{Caa05} \cite{Sue13} have attempted to address this last issue by only displaying PCPs for subsets of the dimensions that are highly correlated.
Given the example dataset and situation specified above, these techniques would construct two PCPs: one displaying $a$, $b$, and $c$, and the other displaying $b$ and $d$.
This enables users to view correlations between $b$ and all of the other dimensions.
However, adjusting the optimal visualization parameters (subset of dimensions to display) is a challenging issue that often requires interactive input by an expert user.
We hypothesize that new high-dimensional data visualization capabilities can be enabled by combining multivariate visualization techniques (e.g. PCPs) with an innovative coupling to interactive dimension selection techniques \cite{Elm08} \cite{Tur11} \cite{Yua13} \cite{Zha15}.

\begin{figure}[t]
\centering
\includegraphics[width=\columnwidth]{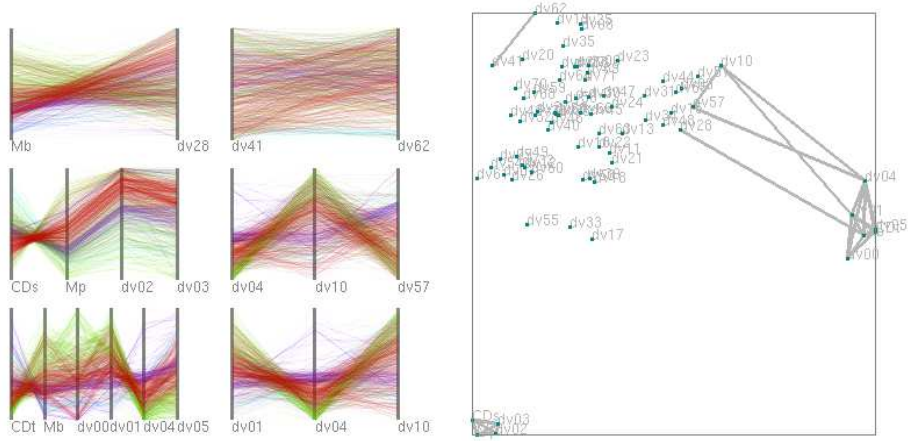}
\caption{A snapshot of our proposed technique.
The left side of the drawing area shows low-dimensional PCPs, and the right side shows a dimension graph.  Distances between arbitrary pairs of dimensions are calculated as a preprocessing.  Groups of close dimensions are extracted and displayed as PCPs. The dimension graph displays how the groups of dimensions are constructed. }
\label{fig:Snapshot0}
\end{figure}

This paper presents a new technique for high-dimensional data visualization in which a set of low-dimensional PCPs are interactively constructed. 
Figure \ref{fig:Snapshot0} shows a snapshot of our proposed technique, which features two visualization components: \emph{low-dimensional PCPs} and a \emph{dimension graph}.
The low-dimensional PCPs display the values of the selected dimensions.
We provide two approaches to interactive dimension selection, derived from the correlation between numeric dimensions or from data mined association rules between numeric dimensions and categorical labels.
The dimension graph displays the relationships between the numeric dimensions and allows interactive dimension selection.
This representation is visually similar to correlation map proposed by Zhang et al. \cite{Zha15}; however, our technique differs in the application of the dimension graph where we enabled simultaneous extraction of a set of low-dimensional subspaces with a simple threshold adjustment operation, thus offering an interactive mechanism that is more intuitive, when compared with existing techniques.
We allow duplication in the selection of dimensions and the consequent display across multiple low-dimensional PCPs. This enables visualization of relationships between a particular dimension and three or more other dimensions, which can be often difficult to understand when all dimensions are shown in a single PCP. At the same time, our method flexibly saves display space because many unnecessary dimensions can be removed from the visualization results.


The rest of this paper is organized as follows. We introduce related work in Section 2. In Section 3, we present the framework that describes our new high-dimensional visualization technique.
We present an experimental validation of our technique in Section 4 and draw conclusions from the outcomes in Section 5.

\section{Related Work}
In this section, we survey PCPs and interactive dimension selection techniques. Our visualization technique builds upon a combination of these approaches.

\subsection{Parallel Coordinate Plots}
As mentioned previously, PCPs \cite{Ins90} display high-dimensional datasets as polylines intersecting with parallel axes.
The improvement of PCPs is a very active research topic and one of the well-known challenges in this domain is that of polyline cluttering, i.e., a reduction of line crossings and overlaps for visual comprehensibility. 
Several techniques have attempted to improve the comprehensibility of the results obtained by PCPs by applying clustering or sampling of the polylines \cite{Fua99} \cite{Joh05} \cite{Ros12} \cite{Zho08}.
In addition, the effectiveness of PCPs are highly dependent on the order of the dimensions and various dimension ordering techniques have recently been proposed to address this issue  \cite{Ferdosi11} \cite{Heinrich12} \cite{Pen04} \cite{Zha12}.
The last major challenge is the difficulty in representing all correlations in one display space, especially when a particular dimension is strongly correlated with many other dimensions.
In these circumstances, PCPs can represent only a subset of all possible relationships between the dimensions.

\subsection{Dimension Selection for High-Dimensional Data Visualization}

When a multi-dimensional dataset contains a very large number of dimensions, existing visualization techniques (e.g., PCPs or SPs) may need very large display spaces to represent them completely.
This problem can be solved by dividing the high-dimensional data space into smaller subsets.
Ten Caat et al. applied multiple PCPs to represent time-varying multidimensional data \cite{Caa05}.
Suematsu et al. \cite{Sue13} also converted high-dimensional datasets into low-dimensional subsets and visualized these subsets using multiple PCPs arranged on display spaces based upon their similarity and correlation.
Using similar ideas, Zheng et al.~\cite{Zhe15} selected SPs based upon the meaningfulness of the dimensions being displayed and adjusted their layout based upon their similarity.
Claessen et al. \cite{Cla11} presented a technique to visualize high-dimensional datasets by selecting sets of low-dimensional subspaces and representing them as a combination of PCPs and SPs.
These techniques provide static results by the decomposition of high-dimensional spaces into multiple pre-selected low-dimensional spaces.  However, the lack of any interactive mechanisms to select the sets of dimensions denies domain experts the ability to use their prior experience and knowledge to specify rules about the data.

Several studies have demonstrated that subsets or subspaces of high-dimensional spaces can be effectively visualized by a user's interactive selection of the subspace (subset of dimensions).
Elmqvist et al. \cite{Elm08} presented an interactive mechanism to select pairs of dimensions and to smoothly switch between the SPs, which was effective in preserving users' mental maps of the high-dimensional spaces. 
Lee et al. \cite{Lee13} and Liu et al. \cite{Liu14} applied dimension reduction schemes to interactively select subsets of the high-dimensional data.
Nohno et al. \cite{Noh14} presented a technique to interactively contract highly-correlated dimensions to adjust the number of axes displayed in PCPs.

Several recent studies have applied SPs for the representation of \emph{dimension spaces} in which each dot in the SP represents a single dimension in the space.
Turkay et al. \cite{Tur11} \cite{Tur12} presented a dual SP model to visualize both the items and dimensions spaces.
Similarly, Yuan et al. \cite{Yua13} presented a interactive mechanism to select low-dimensional subspaces on the SP display in which each dot corresponds to a different dimension.
We also represent the relationships among the dimensions in a 2D space; however, our technique applies a graph rather than a SP.

The technique recently proposed by Zhang et al. \cite{Zha15} uses a similar representation to that applied by our technique. They construct a ``correlation map'' in which the dataset dimensions are represented by dots where the connection between the dots is derived from pairwise correlations.
Our technique includes two characteristics that differ fundamentally from the method of Zhang et al. \cite{Zha15}.
Firstly, the dimension graph in our technique is used as an interactive mechanism to simultaneously control the dimensionality of the set of PCPs, thereby allowing the PCPs to act as a visual representation of a set of low-dimensional subspaces; in contrast, users need to find interesting dimensions and select them individually while using the method of Zhang et al.
In addition, our technique uses association rule mining to extract low-dimensional subspaces in contrast with the correlation-based technique used by Zhang et al. \cite{Zha15}.
Our technique can therefore extract complex multi-variate relationships in comparison to pairwise correlations.

\section{Proposed Visualization Technique}

This section describes our new high-dimensional data visualization technique. 

\subsection{Processing flow}

\begin{figure}[th]
\begin{center}
\includegraphics[width=\columnwidth]{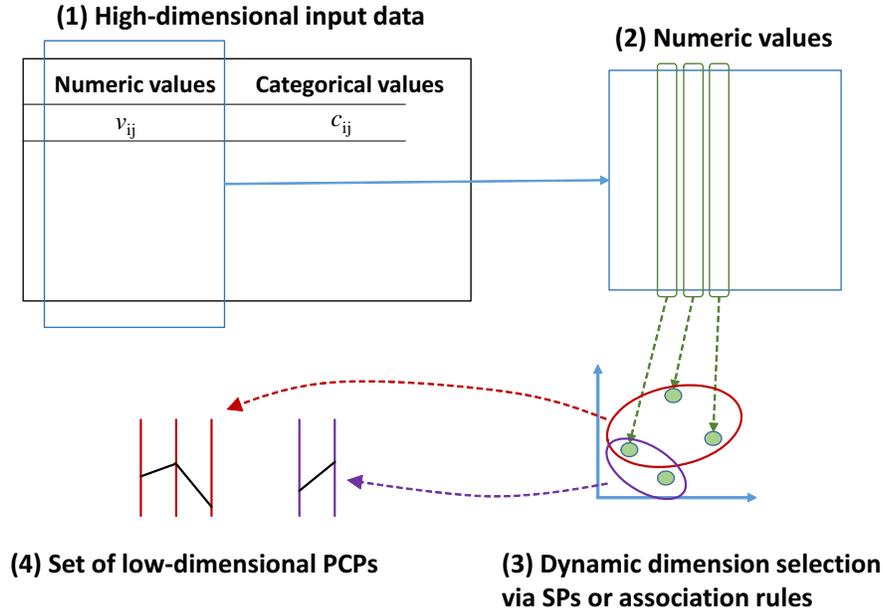}
\end{center}
\caption{Processing flow of our technique. (1) High-dimensional input data. (2) Each dimension of numeric values is treated as a vector, and distances between arbitrary pairs of the set of vectors are calculated. (3) Sets of dimensions are semi-automatically selected. (4) The sets of dimensions are displayed as low-dimensional parallel coordinates. }
\label{fig:Overview1}
\end{figure}

Figure \ref{fig:Overview1} illustrates the processing flow of our technique.
Our technique treats the values of numeric dimensions as vectors, and calculates the distances between arbitrary pairs of dimensions.
It selects the sets of dimensions with an interactive threshold setting.
Our current implementation provides two types of dimension selection schemes.
The first approach is based on distances between arbitrary pairs of dimensions, as described in Section \ref{sec:dimselect1}.
This approach forms a graph by connecting the pairs of dimensions if their distances are smaller than the threshold, and finds cliques of the graph as sets of dimensions.
The second approach is based on concentration or separateness of the labels, as described in Section \ref{sec:dimselect2}.
The approach treats the value of a user-specified categorical dimension as \emph{labels}, and extracts sets of dimensions if particular labels purely concentrate on the particular portions of the axes of PCPs.
The technique allows the user to modify the threshold via a slider widget on the window system, and selects the sets of dimensions and updates the sets of PCPs accordingly.

We formalize the high-dimensional datasets visualized by our technique as follows.
The dataset has $m$ items, which contain $n$-dimensional values, including $n_v$ numeric dimensions and $n_c$ categorical dimensions.
The dataset $D$ and the $i$-th item $a_i$ are described as:
\begin{eqnarray}
 D = \{a_1, ..., a_m \} \nonumber \\
 a_i = \{v_{i1}, ..., v_{in_v}, c_{i1}, ..., c_{in_c} \} \nonumber
\end{eqnarray}
where $v_{ij}$ denotes the $j$-th numeric dimension value of the $i$-th item, and $c_{ik}$ denotes the $k$-th categorical dimension value of the $i$-th item.
The value of $c_{ik}$ should be one of the categorical values $\{C_{k1}, ..., C_{kn_{k}}\}$, where $n_{k}$ denotes the number of possible categorical values for the $k$-th categorical dimension.

\subsection{Dimension Set Selection}

Dimension set selection is a key component for our visualization technique. The requirements for the dimension set selection are strongly motivated by what individual users want to see and how they wish to explore the data. In this paper, we describe two approaches for interactive dimension set selection.

\subsubsection{Distances among dimensions} \label{sec:dimselect1}

It is important to understand correlation or relationships among variables in many high-dimensional data analysis applications.
Thus we provide a scheme to select sets of dimensions in which similar or highly correlated dimensions are grouped into the same set.
This process generates a dimension graph by connecting highly correlated dimensions, and extracts cliques of the dimension graph as sets of highly correlated dimensions.
The dimension graph is visualized by applying a dimensionality reduction scheme.

We treat each numeric dimension as a vector, described as
$\{v_{j1}, v_{j2}, ...v_{jn}\}$ for the $j$-th dimension, where $n$ is the number of items.
We first calculate the distances between all possible pairs of the numeric dimensions.
The distance between the $j$-th and $k$-th numeric dimensions is defined as:
\begin{equation}
d_{jk} = |1.0 - f_c(j,k)|
\label{eq:dist}
\end{equation}
where $f_c(j,k)$ denotes Spearman's rank correlation coefficients.
This definition means that positively or negatively correlated numeric dimensions have similar distances and are thus close.

We then form a graph by connecting pairs of numeric dimensions if their distances $d_{jk}$ are smaller than user-defined threshold $d_{select}$.
Consequently, more connections are generated if the user set larger thresholds.
Next, we extract the set of \emph{cliques}, or complete subgraphs, by performing a maximum clique detection.
Our implementation applies the Bron-Kerbosch algorithm \cite{Bro73}.
The sets of dimensions corresponding to the cliques are displayed as PCPs.

\begin{figure}[th]
\begin{center}
\includegraphics[width=\columnwidth]{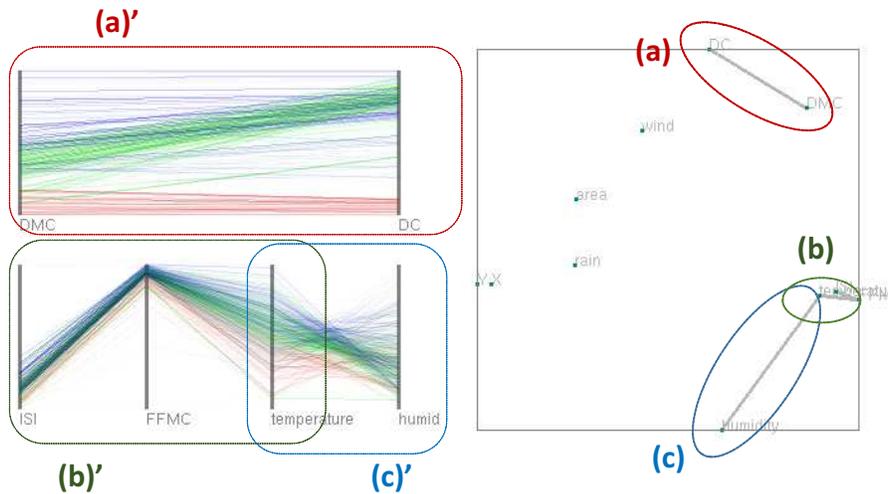}
\end{center}
\caption{Dimension set selection according to distances among dimensions.
(Left) A set of PCPs displayed from the selected sets of dimensions. (Right) Three cliques of the graph constructed by connecting pairs of dimensions if their distances are sufficiently small. The cliques (a) to (c) correspond to (a)' to (c)' in the PCPs.}
\label{fig:DistanceClique}
\end{figure}

Figure \ref{fig:DistanceClique} shows a basic example of dimension set selection result.
The right side of the window displays the dots corresponding to the numeric dimensions and their connections.
We calculate the positions of the dots based on the distances $d_{jk}$ by applying MDS (multi-dimensional scaling)
\footnote{We apply the classical MDS implemented in MDSJ: Java Library for Multidimensional Scaling (Version 0.2). Available at http://www.inf.uni-konstanz.de/algo/software/mdsj/. University of Konstanz, 2009}.
The left side of the window displays the PCPs generated from the dimensions contained in the cliques of the graph.
Our implementation allows two PCPs to be merged if their corresponding cliques share only a single dimension, e.g., PCPs (b)' and (c)' in this figure.

We reorder the numeric dimensions so that well correlated pairs of dimensions are adjacently placed in the PCPs. 
We attempt to minimize the sum of $d_{jk}$ between dimensions adjacently drawn in the PCPs, applying an approximate solution of the traveling salesman problem \cite{Zha12}.

\subsubsection{Concentration and separateness of labels} \label{sec:dimselect2}

Categorical dimensions in datasets can be used as labels that indicate particular characteristics of data elements. 
The correlations or relationships between these labels and the numeric values in other dimensions quantify important information in many high-dimensional data analysis and exploration applications.
We therefore also also provide a scheme to select sets of dimensions where particular labels are either concentrated or distinctly separated. 
This scheme firstly applies an association rule mining method to extract well related combinations of labels and ranges of numeric dimensions, and visualizes them as a set of PCPs.

Note that the distance-based dimension selection (Section~\ref{sec:dimselect1}) was a general approach to select dimensions and therefore required the calculation of the correlation between all pairs of features. 
In contrast, the presence of labels through categorical dimensions allows the opportunity to reduce the dimensional space to only those dimensions that have strong relationships with the labels.

To extract association rules between labels and ranges of numeric dimensions, we first divide each range of numeric dimensions into a same number of subranges.
Let the range of one subrange of the $j$-th numeric dimension $div_j$, and the minimum and maximum values be given by $v_{jmin}$ and $v_{jmax}$.
Here, the $k$-th subrange of the $j$-th numeric dimension is described as follows:
\begin{equation}
V_{jk} = [\frac{kdiv_j}{v_{jmax} - v_{jmin}}, \frac{(k+1)div_j}{v_{jmax} - v_{jmin}}] 
\end{equation}

We count the number of items where the value of the $j$-th numeric dimension is in the $k$-th subrange.
We then apply a quantitative association rule mining algorithm \cite{Fuk96} \cite{Sri96}
\footnote{We used our own implementation of association rule mining instead of any publicly available libraries.}
to discover the rules given by:
\begin{equation}
L \to V_{jk}
\label{eq:rule1}
\end{equation}
or
\begin{equation}
V_{jk} \to L
\label{eq:rule2}
\end{equation}
where $L$ is a particular label corresponding to a particular value of a categorical dimension of the input dataset.
The first rule (Equation~\ref{eq:rule1}) denotes the case where the ranges of particular numeric dimensions are predicted from labels of items.  In other words, this rule describes the subspaces where specific labels are concentrated.
The second rule (Equation~\ref{eq:rule2}) denotes the case where labels of items are predicted based upon values of particular numeric dimensions.  In other words, this rule describes the subspaces that separate labels.
We extract the sets of dimensions that correspond to these discovered rules and visualize them as sets of low-dimensional PCPs.
Users can interactively adjust the thresholds of support ($t_{sup}$) and confidence ($t_{con}$), which are commonly used as criteria of association rule mining.

\begin{figure}[th]
\begin{center}
\includegraphics[width=\columnwidth]{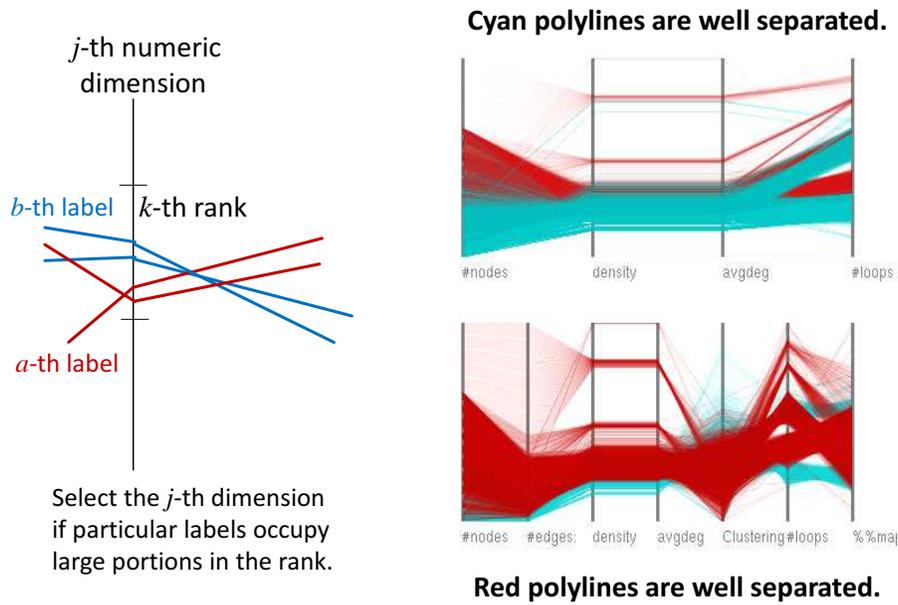}
\end{center}
\caption{Dimension set selection according to separateness of labels.
(Left) Numeric dimensions are divided into multiple subranges. Items drawn as polylines are colored according to values of the user-specified categorical dimension. For each color, the number of polylines intersecting a particular subrange are counted.
(Right) An example of PCPs comprising of numeric dimensions where particular colors of polylines are well-separated.}
\label{fig:LabelSeparateness}
\end{figure}

Figure \ref{fig:LabelSeparateness} illustrates this process applying Equation~\ref{eq:rule2}. It shows an example of PCPs of numeric dimensions where different poly lines (indicated by color) are well-separated.
We define the order of dimensions by using an approximate solution for the traveling salesman as described in Section \ref{sec:dimselect1}.

\subsection{Dimension Sampling} \label{sec:sampling}

\begin{figure}[th]
\begin{center}
\includegraphics[width=7.5cm]{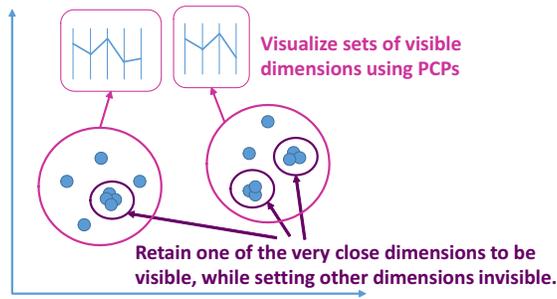}
\end{center}
\caption{Illustration of dimension set selection and sampling.  Our technique retains only one of the very close dimensions to be visualized by PCPs. }
\label{fig:Sampling}
\end{figure}

It is often important to render appropriate number of nodes in a limited display space to comprehensively represent graphs.
Meanwhile, many multidimensional datasets consist of groups of very similar dimensions where it is not necessary to observe every dimension.
Our technique supports a dimension sampling scheme to render dimensions invisible when they are sufficiently close to one or more visible dimensions.
For this process, we again use Equation~\ref{eq:dist} to calculate the distance between dimensions.
We extract pairs of dimensions where these distances are smaller than a user-defined threshold $d_{remove}$, where $d_{remove} < d_{select}$, and randomly select one of these dimensions to make it invisible.  The visible dimensions are then visualized by PCPs. 
Figure \ref{fig:Sampling} illustrates how dimensions are made invisible.

\subsection{Interaction}
\label{sec:interact}

\begin{figure}[th]
\begin{center}
\includegraphics[width=\columnwidth]{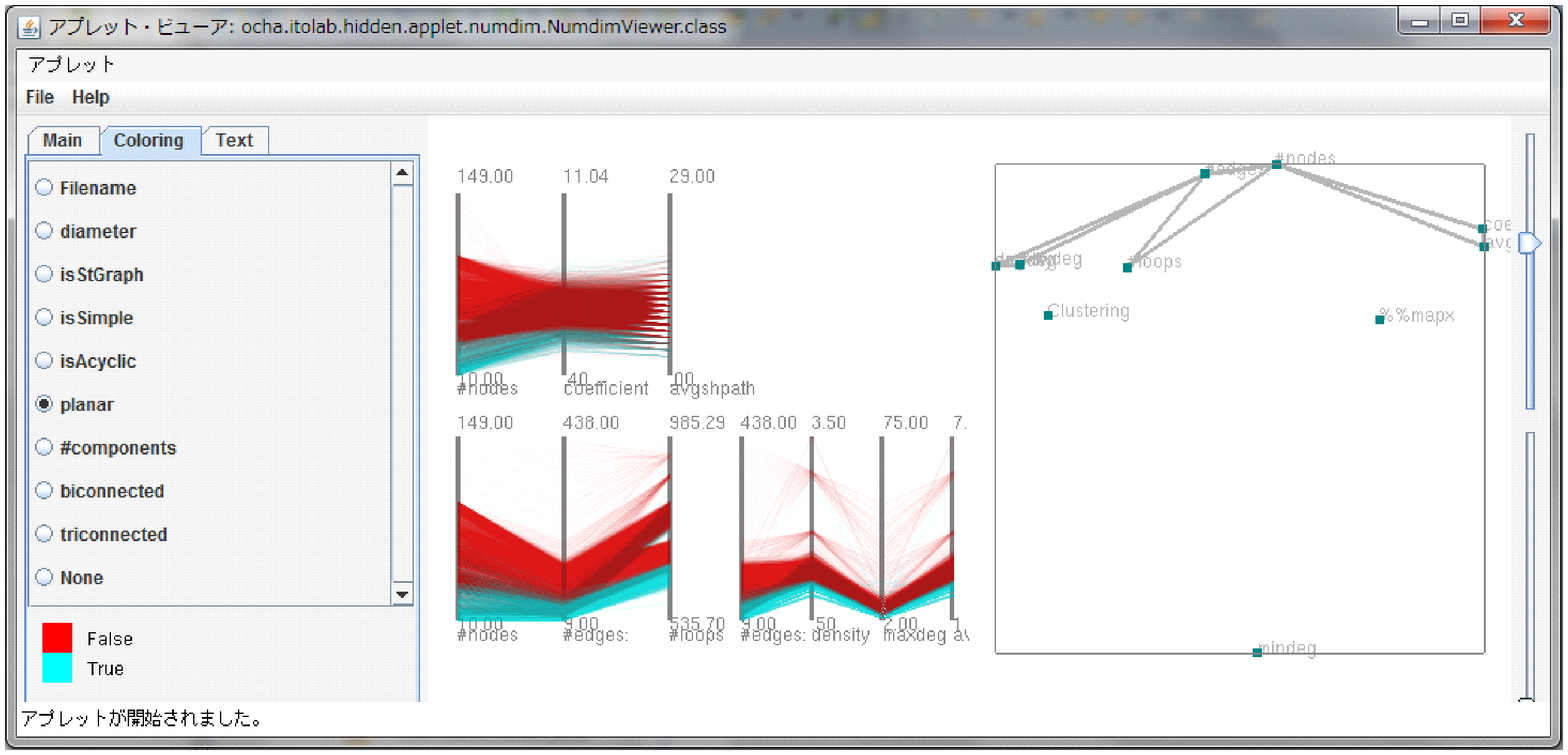}
\end{center}
\caption{Snapshot of out implementation with user interface widgets. The left side of the window features the radio buttons to select a categorical dimensions.  The center of the window draws PCPs.  The right side of the window represents distances among dimensions.}
\label{fig:Snapshot1}
\end{figure}

Figure \ref{fig:Snapshot1} shows a snapshot of our visualization technique.
The upper-left side of the window features the radio buttons to select one of the categorical dimensions.
After selection, the lower-left side of the window displays a list of categorical values and their associated colors, to assist users to recognize the numerical distributions with the categorical values.
In this snapshot, the selected categorical dimension contains two values, ``False'' associated to red, and ``True'' associated to cyan.

The right side of the drawing area displays the dimension graph. Each vertex (dot) corresponds to a numeric dimension. Edges (connections) between pairs of numeric dimensions are formed if their distances are smaller than the user-specified threshold $d_{select}$.
The threshold can be smoothly controlled by users through slider widgets featured at the side of the window.
The slider widgets are also used to control the support and confidence thresholds while applying the association rule mining.
Users may also draw rectangles (using mice or other pointing devices) to specify the dimensions that should be forcibly included or excluded from the selected sets of dimensions.

The left side of the drawing area displays the sets of PCPs.
The PCPs are updated when users select different categorical values using the radio buttons, adjust the thresholds $d_{select}$ or $d_{remove}$ with the slider widgets, or draw rectangles within the drawing area.

\subsection{Rendering by PCPs}
\label{sec:pcprender}

The visualization of the low-dimensional subspace occurs after the interactive selection of numeric dimensions described in Section~\ref{sec:interact}.
While we use PCPs for our visualization, other techniques can also be applied, e.g., SPs can be used instead of PCPs by limiting the number of dimensions in each of the selected subsets.
In our visualization, the axes of the PCPs are ordered as described in Section \ref{sec:dimselect1}.
Polylines are then drawn; the transparency of the polylines in our current implementation can be controlled with a slider widget.

Polyline coloring is powerful visual tool for assisting users to distinguish polylines in PCPs. 
In our implementation, the colors are defined according to the values of the user-selected categorical dimension. 
However, there are many datasets without any categorical dimensions thus precluding color assignment based on categories. 
In these cases, we therefore divide the items into several groups using $k$-means clustering \cite{Har79} and assign the colors accordingly; the number of clusters can be specified by the user.

\section{Experiments}

We performed two case studies with our visualization technique: optimization of airplane wing shape design, and knowledge mining between features and annotations of medical images. Each case study used a different form of dimension set selection: distances among dimensions for the airplane wing shape case study and the concentration and separateness of labels (association rules) for the medical imaging case study. Both of these problem domains have high-dimensional attributes and hence it is difficult to apply existing low-dimensional approaches such as those mentioned in Section 2. 
We implemented our technique using Java Development Kit (JDK) 1.7.0, and executed it on a Lenovo ThinkPad T450 (2.60MHz Dual Core, RAM 8GB) with Windows 7 (64 bit).

\subsection{Case Study 1: Optimization for Airplane Wing Shape Design} \label{sec:case1}

Our first case study examined our visualization technique in its application to analyzing the variables used for the optimization of airplane wing shape design \cite{Oba00} \cite{Sas02}.
The dataset consists of 72 design variables of whole wing shapes and 4 objective functions obtained from fluid dynamics simulations.
The dataset contains 776 Pareto optimal solutions obtained by a multi-objective genetic algorithm.
These solutions satisfy Pareto efficiency, a state of allocation of resources in which it is impossible to make any one item better off without making at least one item worse off.
We visualized this dataset as 776 items of a 76 dimensional dataset. Note that this dataset does not contain any categorical dimensions; we used the clustering technique described in Section~\ref{sec:pcprender} to assign four colors to the polylines of the PCPs.

One of the motivations for visualizing this dataset is to observe and understand the trade-offs among the objective functions; such trade-offs can be found in many multi-objective optimization problems.
We believe that visualization can contribute to the careful analysis of the distribution of the objective functions thereby enabling a better subjective selection of Pareto solutions and the design of better optimization processes.
Another motivation is to discover unknown relationships among all the variables, not limited to a subset ($dv_{00}$ to $dv_{05}$ as described below).
Such discoveries will be useful in narrowing down the variables to be optimized during the design of better wing shapes.

The design variables in the dataset are given by the range from $dv_{00}$ to $dv_{71}$.
It is well-known in aerospace design community that the following six variables were most important for the optimum solution discovery:
\begin{itemize}
\item $dv_{00}$, $dv_{01}$: Span lengths of the inboard/outboard wing panels.
\item $dv_{02}$, $dv_{03}$: Leading-edge sweep angles.
\item $dv_{04}$, $dv_{05}$: Root-side chord lengths.
\end{itemize}
Other design variables included the following:
\begin{itemize}
 \item $dv_{06}$ to $dv_{25}$: Variables to define the inner surface connecting corresponding points on upper and lower surfaces of the wing, to design the warping of the wing.
 \item $dv_{26}$ to $dv_{32}$: Variables to design the twist of the wing.
 \item $dv_{33}$ to $dv_{71}$: Variables to design the thickness of the wing.
\end{itemize}
The objective functions are as follows:
\begin{itemize}
\item $CD_t$: Drag coefficient during transonic cruise.
\item $CD_s$: Drag coefficient during supersonic cruise.
\item $M_b$: Bending moment at the wing root during supersonic cruise.
\item $M_p$: Pitching moment during supersonic cruise.
\end{itemize}

\begin{figure}[th]
\begin{center}
\includegraphics[width=\columnwidth]{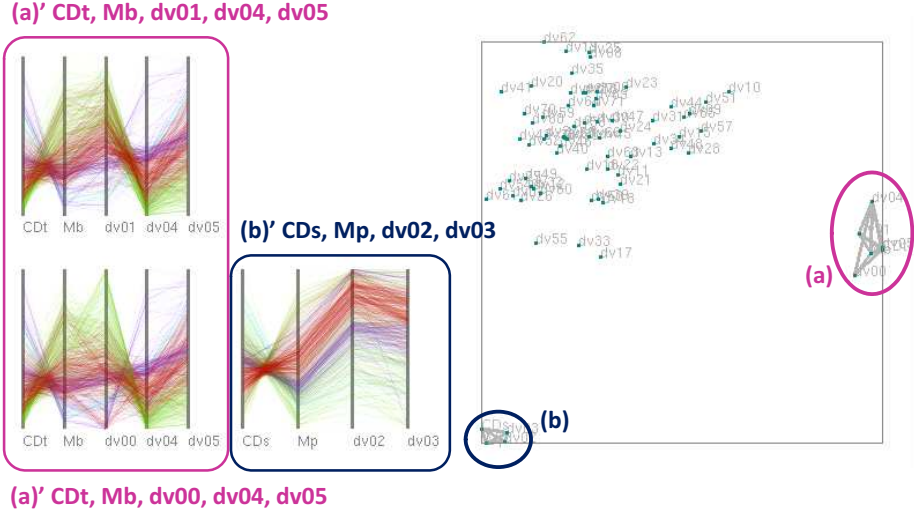}
\end{center}
\caption{Pareto solutions of multi-objective optimization for airplane wing design (1). The six design variables $dv_{00}$ to $dv_{05}$ were known to be dominant for the design optimization, and these variables are strongly correlated with the four objective functions. A clique (a) is represented by PCPs (a)', while another clique (b) is represented as another PCP (b)'.}
\label{fig:MultiOpt1}
\end{figure}

We first observed the positions of dots in the right side of the window as shown in Figure \ref{fig:MultiOpt1}.
We found that six dots, corresponding to the four design variables $dv_{00}$, $dv_{01}$, $dv_{04}$, and $dv_{05}$, and the objective functions $CD_t$ and $M_b$, were closely placed in the right end as shown in Figure \ref{fig:MultiOpt1}(a).
We also found that four dots, corresponding to the design variables $dv_{02}$ and $dv_{03}$, and the objective functions $CD_s$ and $M_p$, were similarly concentrated as shown in Figure \ref{fig:MultiOpt1}(b). 

We then interactively adjusted the threshold $d_{select}$ and observed the PCPs to determine the correlations among the dimensions.
As a result, we discovered correlations between pairs of the variables, including negative correlation between $CD_t$ and $M_b$ as in Figure \ref{fig:MultiOpt1}(a)'.
This negative correlation denotes a typical trade-off between the two objective functions.
Meanwhile, $dv_{00}$ and $dv_{01}$ had negative correlations with $dv_{04}$, while $dv_{04}$ had a positive correlation with $dv_{05}$.
Figure \ref{fig:MultiOpt1}(b)' shows that another pair of objective functions $CD_a$ and $M_p$ has a trade-off.
It also shows that the positive correlation between $dv_{02}$ and $dv_{03}$ brings Pareto solutions. 

\begin{figure}[th]
\begin{center}
\includegraphics[width=\columnwidth]{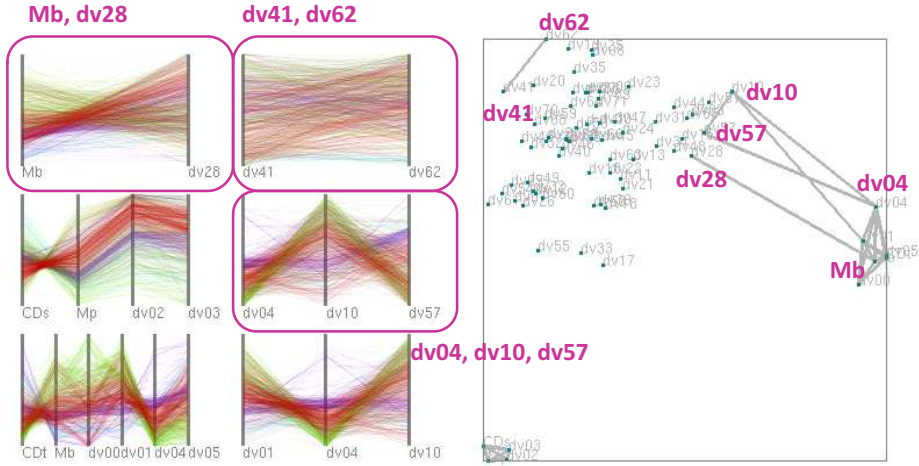}
\end{center}
\caption{Pareto solutions of multi-objective optimization for airplane wing design (2).
We identified many unknown relationships among design variables and objective functions by interactively adjusting the threshold parameter $d_{select}$.}
\label{fig:MultiOpt2}
\end{figure}

Figure \ref{fig:MultiOpt1} also demonstrates that our technique effectively visualizes the subset of dimensions that are strongly correlated.
However, the correlation visualized in this figure was already a well-known property of airplane wing design.
It is perhaps a more interesting and important problem to use our visualization technique to uncover unknown relationships among the variables. 
We therefore visualized more relationships by interactively adjusting the threshold parameter $d_{select}$; the resulting visualization is shown in Figure \ref{fig:MultiOpt2}.
We found negative correlations between $M_b$ and $dv_{28}$, $dv_{41}$, and $dv_{62}$, $dv_{04}$ and $dv_{10}$, and $dv_{10}$ and $dv_{57}$.
Here, $dv_{10}$ is one of the design variables to define the shape of curved surface connecting corresponding positions on upper and lower surfaces of the wing, which controls the crook of the wing;
$dv_{28}$ is one of the variables to design the twist of the wing; and, $dv_{41}$ and $dv_{62}$ control the thickness of the wing. 
The dataset owner, an expert in aerospace fluid simulations, communicated that he did not know this result and it might bring new knowledge in the field of airplane design.

Several related visualization techniques \cite{Tur11} \cite{Yua13} apply dimension-reduced scatterplots for representation of low-dimensional subspaces.
This representation is suitable to find similar simulation results; on the other hand, our technique is more convenient for carefully observing how dimensions correlate to each other.
The correlation map \cite{Zha15} is also useful for observing the relationships among the dimension; however, users need to find interesting dimensions and select them individually on the correlation map.  Our technique demonstrated its ability to simultaneously extract all sets of well-correlated low-dimensional subspaces with a simple threshold adjustment operation.

\begin{figure}[th]
\begin{center}
\includegraphics[width=\columnwidth]{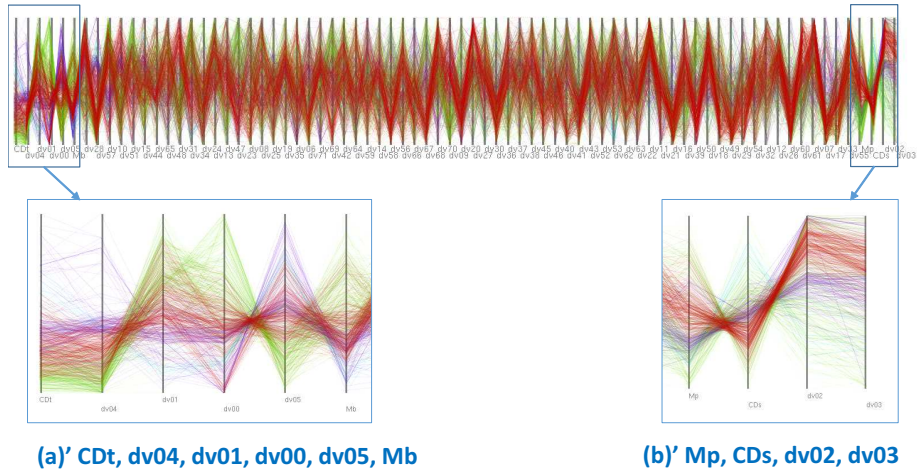}
\end{center}
\caption{Visualization of Pareto solutions using common PCP.}
\label{fig:MultiOptOnePCP}
\end{figure}

Figure \ref{fig:MultiOptOnePCP} shows an example of visualization of Pareto solutions using common PCP.
It is hard to recognize relationships between adjacent dimensions unless we have very horizontally wide display spaces.
Dimension selection is useful to compactly visualize sets of well-correlated dimensions.
Also, it is hard to analyze relationship between a particular dimension and each of more than two other dimensions using regular PCPs, because they are useful mainly while observing relations between adjacent dimensions.
It is more useful to observe complex relationships between such dimensions using our technique, which allows to duplicate a single dimension to appear in multiple PCPs.

\subsection{Case Study 2: Features and Annotations of Medical Images} \label{sec:case2}

Medical data informatics is an increasingly important research area in healthcare~\cite{KumA}. A key challenge in this area is the analysis and interpretation of multidimensional data for retrieval and classification applications. Most existing approaches rely on automatic ``black-box'' approaches that use machine-learning to select the most discriminative features, with limited opportunity for input or verification by domain experts, e.g., artificial neural networks to predict the characteristics of lung nodules~\cite{Kim10}, decision tree committees for medical case retrieval~\cite{Quellec10}, etc. However, it has been shown that black-box learning can train models that do not have meaningful reasons for their predictions. For example, computer-selected chemicals did not perform as predicted in real-world physical experiments~\cite{Gabel14}. Transparency is therefore an important aspect, especially for medical informatics, as it engenders trust in the system by allowing human experts the opportunity to verify and validate the choices that have been automatically made~\cite{Solt09}. Visualization introduces an element of transparency to the feature selection process and provides deeper insights into the correlations and associations between different dimensions. Visualizations can improve verifiability by showing that the combinations of inter-related dimensions correspond to a particular clinical outcome. Furthermore, visualizations also offer the opportunity for an exploration of the dataset that be used to identify outliers and new unknown patterns~\cite{KumB}. 

Clinical experts usually interpret volumetric (3D) medical images through multi-planar views that show the images from the three standard orientations or planes; each plane shows similar or related information from different points of view~\cite{Cho05}. In the literature, many medical image informatics systems use features extracted from the axial ({\emph x-y}) viewing plane only~\cite{Kim10}. In this case study, we will visualize the image features extracted from a large dataset of volumetric medical images to validate whether axial features are sufficient to predict diagnostic outcomes. 

We obtained 933 computed tomography (CT) images of lung cancer patients from the Lung Image Database Consortium and Image Database Resource Initiative (LIDC/IDRI)~\cite{Arm11}\footnote{The LIDC/IDRI dataset source: \url{http://cancerimagingarchive.net/}}.
The dataset also included four annotations assigned by clinical experts; two annotations described the diagnostic outcomes of the images and the other two indicated the method used to determine the diagnosis. Our visualizations  plotted the image features (dimensions) that were related to the following annotations:
\begin{itemize}
 \item \emph{Nodule-Level Diagnosis}:
   Diagnosis assigned to the first lung growth or lesion.
   (0 = unknown, 1 = benign or non-malignant disease,
   2 = malignant primary lung cancer, 3 = malignant metastatic diseases.)
 \item \emph{Method for Nodule-Level Diagnosis}:
   The clinical method used to determine the diagnosis for the first/primary nodule.
   (0 = unknown, 
   1 = review of radiological images to show 2 years of stable nodule
   2 = biopsy, 3 = surgical resection, 4 = progression or response.)
\end{itemize}

We used automatic image processing techniques to calculate well-established image features from the nodules in each image~\cite{Kim10}. For each image, we extracted the same set of 316 dimensional features from the three standard viewing planes (axial: {\emph x-y} plane, coronal: {\emph x-z} plane, sagittal: {\emph y-z} plane), resulting in 948 dimensions in total.
Our method is capable of operating in the cases where the number of dimensions are larger than the number of images.

While image features from the three standard planes enables consideration of the complete information about the image it may also introduce redundant information through highly correlated features extracted from different planes. The features we extracted were categorized as follows:
\begin{itemize}
 \item Nodule size and shape properties (e.g., area, convex hull, circularity, etc.).
 \item Pixel intensities inside the nodule (e.g., range, mean, standard deviation, etc.).
 \item Texture of the nodule, described using grey-level co-occurrence (GLC) features (e.g., entropy, contrast, etc.) and Gabor filter responses.
\end{itemize}

\begin{figure}[th]
\begin{center}
\includegraphics[width=\columnwidth]{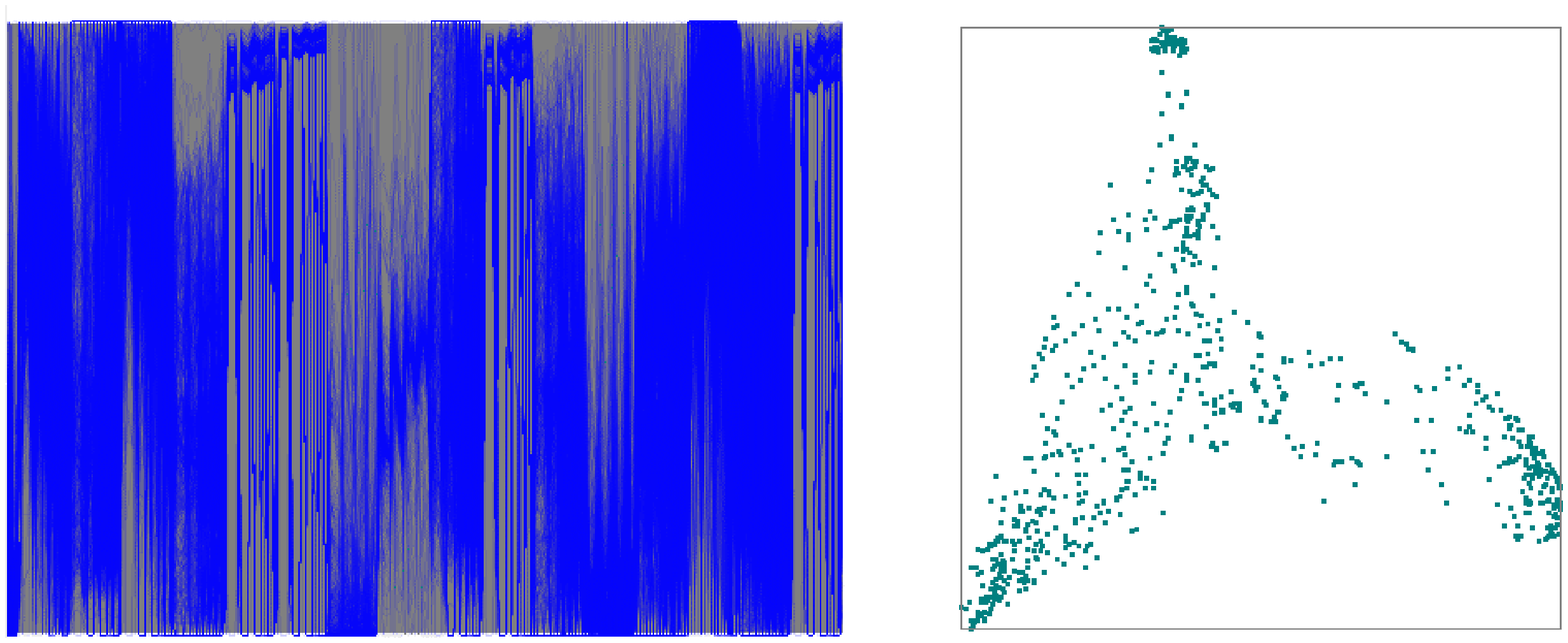}
\includegraphics[width=\columnwidth]{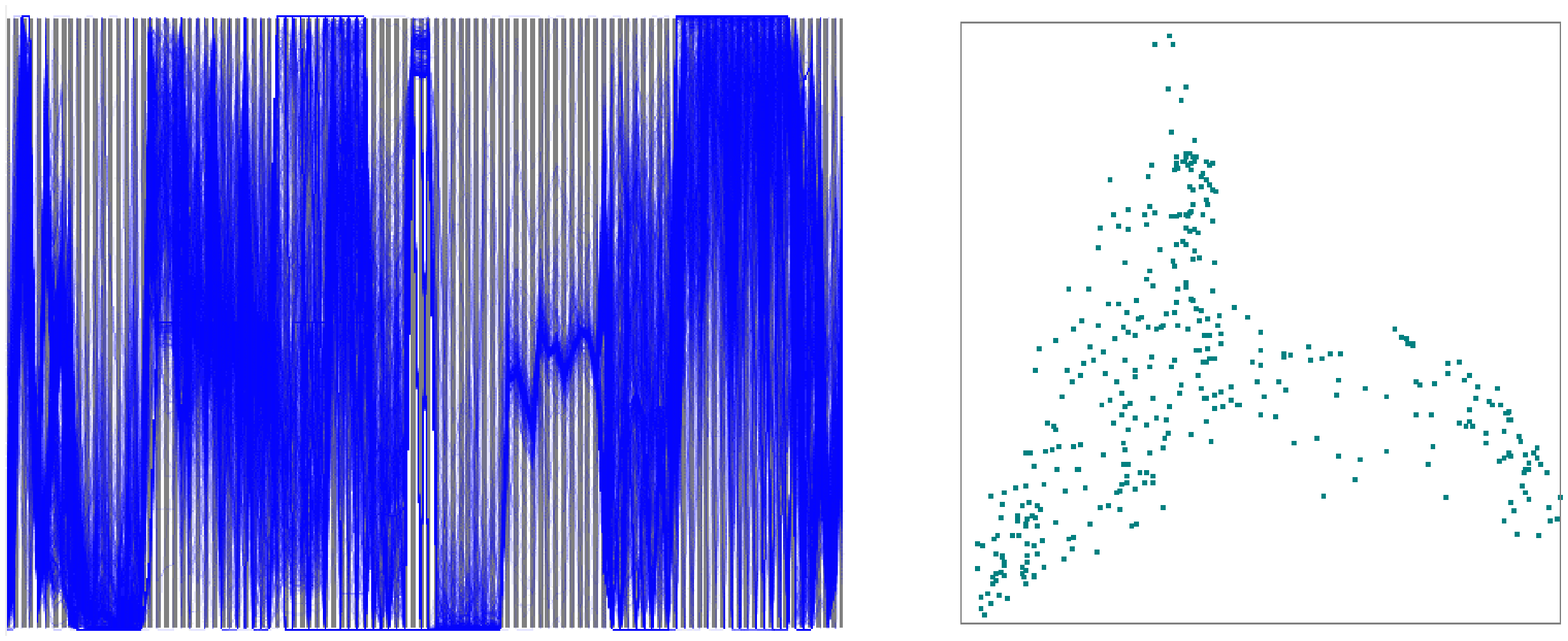}
\end{center}
\caption{PCP and dimension graph of the multidimensional feature values derived from the LIDC images. (Upper) Visualization of all the 948 image features. (Lower) Sampling of 364 features.}
\label{fig:LIDC1-2}
\end{figure}

Figure \ref{fig:LIDC1-2} shows an initial visualization of the dataset.
The high degree of inter-relationships between the image features means that it is quite difficult to understand numeric distribution of individual dimensions by a simple observation of the PCPs.
This remains true even after we visualized a subset of the 364 features selected by the distance among dimensions sampling (Section \ref{sec:sampling}).

We therefore visualized a separate set of PCPs for the subspace of image features that were associated with each of the labels within the annotations. We interactively determined the low-dimensional subspace as follows. We first selected an annotation and data mined the rules that indicated which numeric dimensions were responsible for separation or concentration of the labels within the annotation (see Section~\ref{sec:dimselect2}). We then adjusted the thresholds $t_{sup}$ and $t_{con}$ to vary the strength of the concentration or separation until visually clear PCPs were discovered. 

\begin{figure}[th]
\begin{center}
\includegraphics[width=9cm]{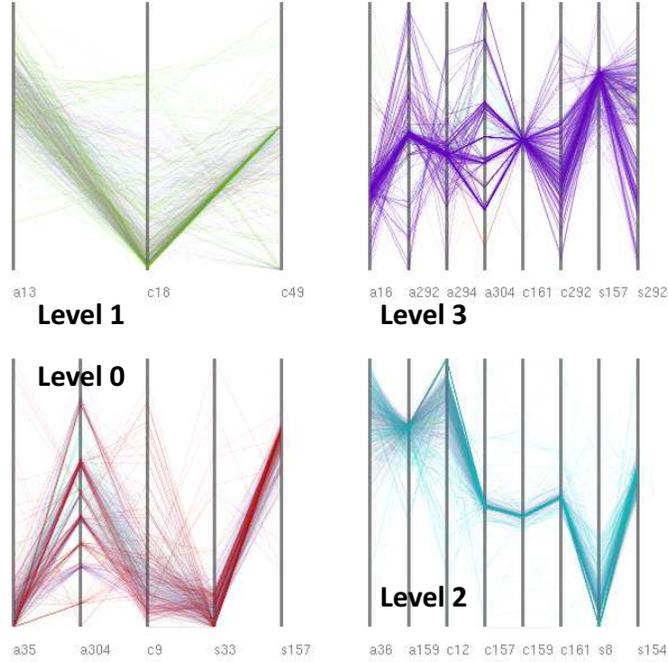}
\end{center}
\caption{Dimension set selection based on separateness of labels. Polylines are labeled and colored according to the \emph{Nodule-Level Diagnosis} annotation. The four PCPs correspond to the four levels of the Nodule-Level Diagnosis. }
\label{fig:LIDC3a}
\end{figure}

\begin{figure}[th]
\begin{center}
\includegraphics[width=9cm]{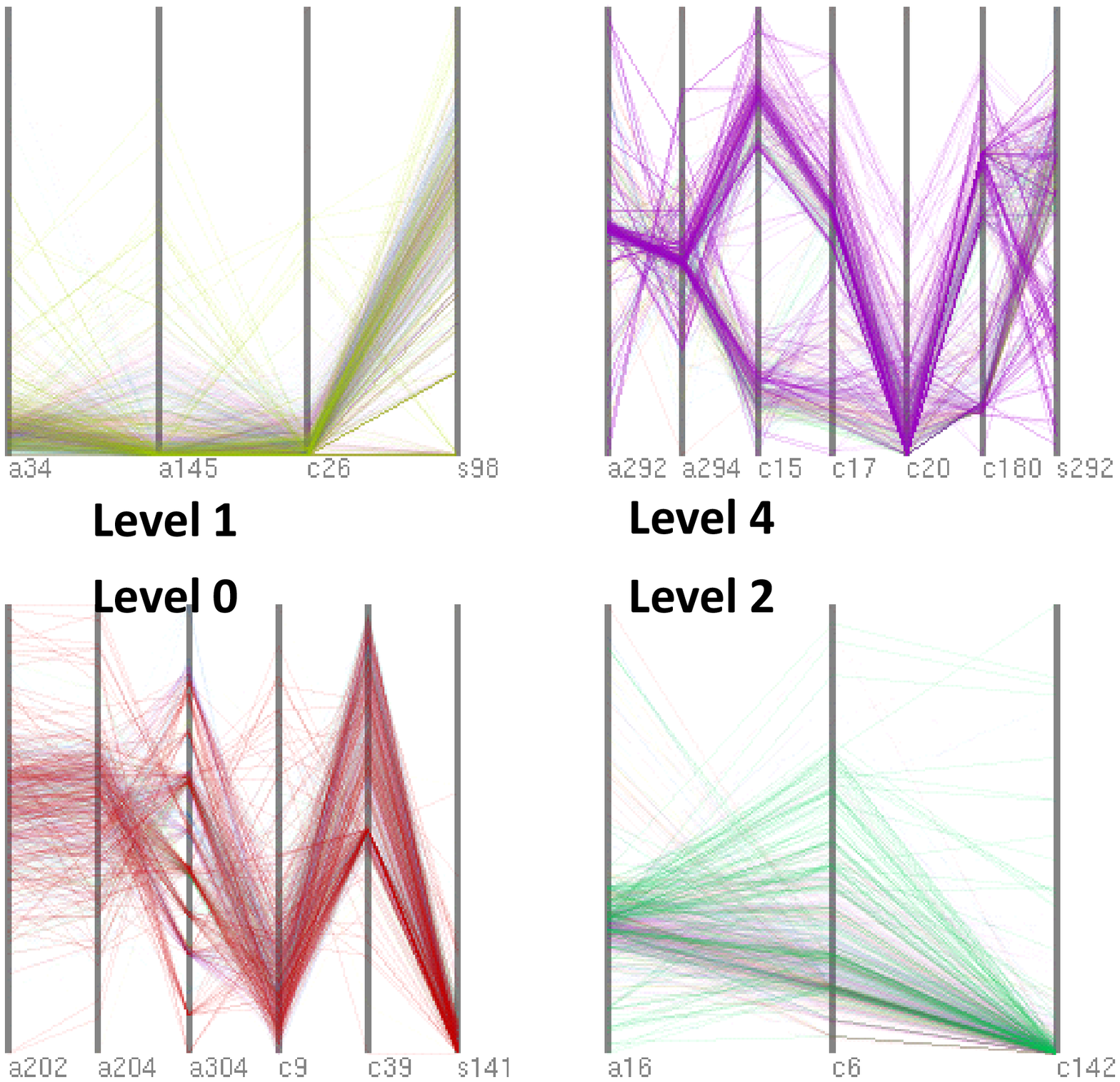}
\end{center}
\caption{Dimension set selection based on separateness of labels. Polylines are labeled and colored according to the \emph{Method of Nodule-Level Diagnosis} annotation.  The \emph{Method of Nodule-Level Diagnosis} has five levels, but this result shows four PCPs because no rules with sufficiently high support could be mined for level 3.}
\label{fig:LIDC3b}
\end{figure}

The resultant visualizations are depicted in Figures \ref{fig:LIDC3a} and \ref{fig:LIDC3b}.
The axes of these PCPs represented the most meaningful image features for each different label.  
The IDs of individual image features are indicated in the figures, and their names are shown in Tables \ref{tab:Annotation3} and \ref{tab:Annotation4}.
Figure~\ref{fig:LIDC3a} shows the most meaningful features for the annotations of \emph{Nodule-Level Diagnosis}, in which the polylines for label ``0'' are drawn in red, ``1'' in green, ``2'' in cyan, and ``3'' in purple; the names of these features are summarized in Table~\ref{tab:Annotation3}.
Figure~\ref{fig:LIDC3b} shows the most meaningful features for the annotations of \emph{Method for Nodule-Level Diagnosis}, in which polylines for label ``0'' are drawn in red, ``1'' in green, ``2'' in cyan, and ``4'' in purple; these features are summarized in~Table \ref{tab:Annotation4}. Note that no discriminatory features were found for label ``3'' of \emph{Method for Nodule-Level Diagnosis}, indicating that perhaps new image features need to be developed to account for diagnosis after `surgical resection'. Visually, adjacent axes connected by densely placed parallel polylines indicate the features that are most important to a particular label.

Both the visualization and Table~\ref{tab:Annotation3} show that while axial features are useful, the introduction of features from the other viewing planes could augment the classification of lung abnormalities. In particular, the thick, parallel polylines for the ``coronal GLC texture cluster shade'' feature in the plots for both primary (label 2) and metastatic (label 3) disease suggest that this feature is useful for distinguishing between benign (labels 0 and 1) and malignant disease (labels 2 and 3). Other features have polylines that are only parallel in one plot and so are useful for separating one label from the others, e.g., ``coronal shape solidity'' and ``sagittal shape roughness'' are useful for separating primary disease (initial site of cancer) indicated by label 2 from metastatic disease (cancer that has spread) indicated by label 3.

Similarly, the visualization and Table~\ref{tab:Annotation4} show that image features from all planes are relevant to different diagnostic methods. Axial intensity and shape axis length features are related to diagnosis by biopsy (label 2) and exhibit many parallel polylines. Similarly, coronal GLC texture contrast, coronal intensity, and sagittal entropy would be useful for emulating radiological review (label 1). 

Our findings indicate that a combination of features from different viewing planes offer better discriminatory power for distinguishing different labels and annotations. This is in contrast to existing work on the same dataset that only uses axial features~\cite{Kim10}. Our visualization technique has thus identified new image features that will be more meaningful to a human reader and this could enable the creation of computer-aided diagnostic tools that emulate different diagnostic processes. For example, it could facilitate the use of optimal multi-planar image features (Table~\ref{tab:Annotation3}) in computer-aided diagnosis systems, thereby allowing these applications to mimic the image interpretation of a clinical expert.


\begin{table}[t]
 \begin{center}
 \caption{List of meaningful features that distinguish labels from ``Nodule-Level Diagnosis'' discovered by interactively adjusting $t_{sup}$ and $t_{con}$.}
  \begin{tabular}{|c|c|l|}
   \hline 
    Label & ID & image feature name (parameters) \\
   \hline
    0
    & a35 & axial contrast (distance: 4, angle: 135) \\
    & a304 & axial Gabor filter standard deviation (orientation: 0.0,  scale: 0.3) \\
    & c9 & coronal shape elongation \\
    & s33 & sagittal GLC texture contrast (distance: 4, angle: 45) \\
    & s157 & sagittal GLC texture cluster shade (distance: 3, angle: 45) \\

    1 
    & a13 & axial shape extent \\
    & c18 & coronal intensity standard deviation \\
    & c49 & coronal GLC texture correlation (distance: 4, angle: 45) \\

    2 
    & a36 & axial GLC texture correlation (distance: 1, angle: 0) \\
    & a159 & axial GLC cluster shade (distance: 3, angle: 135) \\
    & c12 & coronal shape solidity \\
    & c157 & coronal GLC texture cluster shade (distance: 3, angle: 45) \\
    & c159 & coronal GLC cluster shade (distance: 3, angle: 135) \\
    & c161 & coronal GLC cluster shade (distance: 4, angle: 45) \\
    & s8 & sagittal shape roughness \\
    & s154 & sagittal GLC texture cluster shade (distance: 2, angle: 90) \\
    
    3  
    & a16 & axial intensity maximum \\
    & a292 & axial Gabor filter mean (orientation: 0.0, scale: 0.3) \\
    & a294 & axial Gabor filter mean (orientation: 0.0,  scale: 0.5)\\
    & a304 & axial Gabor filter standard deviation (orientation: 0.0, scale: 0.3)\\
    & c161 & coronal GLC texture cluster shade (distance: 4, angle: 45) \\
    & c292 & coronal Gabor filter mean (orientation: 0.0, scale: 0.3) \\
    & s157 & sagittal GLC texture cluster shade (distance: 3, angle: 45) \\
    & s292 & sagittal Gabor filter mean (orientation: 0.0, scale: 0.3) \\
    
   \hline
  \end{tabular}	
 \label{tab:Annotation3}
 \end{center}
\end{table}

\begin{table}[t]
 \begin{center}
 \caption{List of meaningful features that distinguish labels for ``Method of Nodule-Level Diagnosis'' discovered by interactively adjusting $t_{sup}$ and $t_{con}$.}
  \begin{tabular}{|c|c|l|}
   \hline 
    Label & ID & image feature name (parameters) \\
   \hline
   0
   & a202 & axial GLC texture sum of entropy (distance: 2, angle: 90) \\
   & a204 & axial GLC texture sum of entropy (distance: 3, angle: 0) \\
   & a304 & axial Gabor filter standard deviation (orientation: 0.0, scale: 0.3) \\
   & c9 & coronal shape elongation \\
   & c39 & coronal GLC texture correlation (distance: 1, angle: 135) \\
   & s141 & sagittal GLC texture cluster prominence (distance: 3, angle: 45) \\
   
   1
   & a34 & axial GLC texture contrast (distance: 4, angle: 90) \\
   & a145 & axial GLC texture cluster prominence (distance: 4, angle: 45) \\
   & c26 & coronal GLC texture contrast (distance: 2, angle: 90) \\
   & s98 & sagittal GLC texture entropy (distance: 4, angle: 90) \\
   
   2
   & a16 & axial intensity maximum \\
   & c6 & coronal minorAxisLength \\
   & c142 & coronal custerProminence (distance: 3, angle: 90) \\
   
   4
   & a292 & axial Gabor filter mean (orientation: 0.0, scale: 0.3) \\
   & a294 & axial Gabor filter mean (orientation: 0.0, scale: 0.5) \\
   & c15 & coronal intensity minimum \\
   & c17 & coronal intensity mean \\
   & c20 & coronal GLC texture contrast (distance: 1, angle: 0) \\
   & c180 & coronal GLC texture sum of variance (distance: 1, angle: 0) \\
   & s292 & sagittal Gabor filter mean (orientation: 0.0, scale: 0.3) \\

   \hline
  \end{tabular}	
 \label{tab:Annotation4}
 \end{center}
\end{table}

\subsection{Computation Time}

We measured computation time of technical components of our technique as shown in Table \ref{tab:Time}.
The terms ``Dataset 1'' and ``Dataset 2'' correspond to the datasets used for Case Studies 1 and 2, respectively.
``Dimension graph setup'' corresponds to computation time for dimension-to-dimension distance calculation and node placement for dimension graph applying MDS.
``Distance-based selection'' corresponds to computation time for distance-based dimension selection and reordering described in Section 3.2.1.
``Label-based selection'' corresponds to computation time for label-based dimension selection described in Section 3.2.2.
We freely operated slider widgets 20 times, and derived average and maximum computation times for distance-based and label-based selection.
Since Dataset 1 did not contain any labels, computation time of label-based selection was measured only with Dataset 2.

The result shows that dimension graph setup may require large computation time for the calculation of pairwise distances between large numbers of dimensions.
However, it is important to note that this process occurs only once for each dataset. 
In our implementation, this is calculated as an offline pre-process prior to visualization; the distances are saved to a data file for later reuse.

The result also shows that distance-based selection took more computation time compared to label-based selection.
Computation time may be exponential to the number of dimensions in the largest clique for distance-based selection.
As such, the generation of large PCPs may require extensive computation time.
Meanwhile, the computation time for label-based selection is proportional to the number of dimensions, items, and labels.
The acceleration of these processes will empower interactive operations with large datasets.

\begin{table}[t]
 \begin{center}
 \caption{Computation time (msec.)}
  \begin{tabular}{|c|r|r|}
   \hline 
   & Dataset 1 & Dataset 2 \\
   \hline
   Number of dimensions & 76 & 948 \\
   Number of data items & 776 & 933 \\
   \hline
   Dimension graph setup & 8101 & 1498219 \\
   Distance-based selection (ave.) & 109 & 3788 \\
   Distance-based selection (max.) & 603 & 32338\\
   Label-based selection (ave.) & - & 1399 \\
   Label-based selection (max.) & - & 5220 \\
   \hline
  \end{tabular}
 \label{tab:Time}
 \end{center}
\end{table}

\section{Discussion} \label{sec:discuss}
Our technique allowed users to intuitively define and select multiple low-dimensional spaces using an interactive dimension selection mechanisms. Both our case studies discovered new patterns within the data that were not apparent when visualizing the entire high-dimensional space. Thus our case studies showed how an effective visualization of a subspace of high-dimensional data could be constructed by leveraging a user's prior knowledge of the application domain. We suggest that our visualization technique can be adapted and expanded through the modification of its constituent components.

In our current approach, we use classical MDS to place the vertices (dots corresponding to numeric dimensions) of the dimension graph (Section \ref{sec:dimselect1}).
However, under MDS it is possible that there may be a distortion of the distances caused by the projection to a two-dimensional space. 
Thus in the view of the dimension graph it is possible that vertices in close proximity may not represent dimensions that are close.
While MDS was sufficient for our chosen applications, we suggest several other schemes that may be more suited to other applications.
Non-linear dimensionality reduction algorithms (e.g. Isomap) may create better layouts but require more computation time.
Graph layout techniques such as force-directed \cite{Ead84} or stress minimization \cite{Gan13} layouts may also improve the visualization of the dimension graph.

Similarly, alternate methods could also be used to define the order of dimensions in the PCPs. 
In our approach we used an approximate solution for the traveling salesman problem (Section \ref{sec:dimselect1}).
Other approaches could also potentially be used to derive a meaningful ordering of the dimensions, especially when the dataset contains categorical values.
For example, for some applications it would be meaningful to sort the axes according to the quality of the separation of categorical values, i.e., the dimensions that give the best separation appear first and those that give the least clear separation appear last.
Sorting axes according to the confidence and/or support values could also be an effective approach for some datasets.
These approaches may assist users in understanding the importance of the displayed dimensions.

Another alternative adaptation would be the integration of our two interactive dimension selection techniques to enable more focused lower-dimensional PCP visualization. In this adaptation, the category-based rule mining would be used to plot the numeric dimensions as a dimension graph, in which the edges would be derived from the support and confidence of the association rules. Our clique-based dimension selection could then be applied to select sets of mined dimensions.

As future work, we will investigate the use of different graph layout algorithms and alternate dimension reordering approaches. 
Also, we aim to introduce several additional functions to the current visualization including:
\begin{itemize}
 \item Brushing and filtering operations on the PCPs to specify the region of interest, and recalculation of dimension graph, as some related studies \cite{Tur11} \cite{Yua13} implement brushing interaction for both items and dimensions.
 \item Applying various definitions of distances for dimension graph.
 \item Applying various layout algorithms to dimension graph, as discussed earlier in Section \ref{sec:discuss}.
 \item Applying various dimension reordering for PCPs, as discussed earlier in Section \ref{sec:discuss}. 
 \item Improvement of rendering for PCPs, including cluttering avoidance, and coloring of overlapped semitransparent polylines.
\end{itemize}
After implementing these functions, we will investigate application-specific optimizations of our technique through user experiments that evaluate the effectiveness of the technique in different domains with a variety of datasets.
We would also like to integrate our visualization into a data visualization tool for medical imaging informatics~\cite{KumB} as a non ``blackbox'' method of feature selection; this would enable clinical users to conduct an expanded analysis of the relationships between CT image features and patient- or nodule-level diagnosis.
All of these experiments will be supported by detailed analysis of the visualization results by domain experts.

\section{Conclusions}
We presented an interactive technique for the visualization of high-dimensional data spaces.
Our technique displays numeric values of the selected sets of dimensions by a set of PCPs, while the dimension graph component displays the relations between the numeric dimensions and provides interactive dimension selection mechanisms.
Our technique features two types of dimension selection criteria: distances among dimensions, and separateness of labels.  The former is used to visualize highly correlated sets of dimensions while the latter is used to visualize the sets of dimensions which sufficiently separate items with specific labels.
We demonstrated the effectiveness of our technique using two case studies.
We discovered important knowledge about unknown relations among design variables of airplane wing shape and objective functions, 
and about relations between image feature values derived from CT images and nodule-level diagnosis.

\section*{Acknowledgements}
This work was supported in part by ARC grants.

%








\end{document}